\documentclass[11pt]{article}
\usepackage{epsfig}

\newcommand{\BABARPubYear}    {04}

\newcommand{\BABARProcNumber} {025}
\newcommand{\SLACPubNumber} {10507}
\newcommand{\LANLNumber} {0406047}

\input babarsym

\newcommand{\El}{E_{\ell}}
\newcommand{\mX}{m_X}
\renewcommand{\BR}{{\mathcal B}(B\to X_c\ell\nu)}
\newcommand{\Ecut}{E_\mathrm{cut}}
\newcommand{\fbinv}{\mathrm{fb}^{-1}}
\newcommand{\GeV}{\mathrm{GeV}}

\long\def\inst#1{\par\nobreak\kern 4pt\nobreak
    {\it #1}\par\vskip 10pt plus 3pt minus 3pt}

\begin{document}
{\pagestyle{empty}

\begin{flushright}
SLAC-PUB-\SLACPubNumber \\
\babar-PROC-\BABARPubYear/\BABARProcNumber \\
%\babar-PUB-\BABARPubYear/\BABARPubNumber \\
hep-ex/\LANLNumber \\
June, 2004 \\
\end{flushright}

\par\vskip 4cm

% Title of the paper
\begin{center}
\Large \bf
Determination of $\Vcb$ and Related Results from \babar
\end{center}
\bigskip

\begin{center}
\large 
M. Morii\\
Harvard University, Department of Physics\\
17 Oxford Street, Cambridge, MA 02138, USA\\
(for the \lbabar\ Collaboration)
\end{center}
\bigskip \bigskip

% Abstract
\begin{center}
\large \bf Abstract
\end{center}
The CKM matrix element amplitude $\Vcb$ was determined using
the data collected by the BABAR detector.
The partial branching fraction, lepton-energy moments,
and hadron-mass moments were measured in inclusive semileptonic
decays $B\to X_c\ell\nu$.
A global fit to a Heavy-Quark-Expansion calculation allowed
precise determination of $\Vcb$, $m_b$, $m_c$, $\BR$
and four non-perturbative parameters.

\vfill
\begin{center}
Contributed to the Proceedings of MESON 2004\\
8$^{th}$ International Workshop on Meson Production,
Properties and Interaction\\
6/4/2004--6/8/2004, Krakow, Poland
\end{center}

\vspace{1.0cm}
\begin{center}
{\em Stanford Linear Accelerator Center, Stanford University, 
Stanford, CA 94309} \\ \vspace{0.1cm}\hrule\vspace{0.1cm}
Work supported in part by Department of Energy contract DE-AC03-76SF00515.
\end{center}

\section{Introduction}
Accurate determination of $\Vcb$,
the CKM matrix element that governs the $b\to c$ weak transition,
is an important basis for testing the unitarity of the CKM matrix.
Semileptonic decays of the $B$ mesons provide the best access
to this quantity as the leptonic current can be cleanly factored out.
Heavy Quark Expansion (HQE) has become a useful theoretical tool
for calculating the QCD corrections needed for the prediction of
the experimental observables such as the inclusive rate
$\Gamma(B\to X_c\ell\nu)$, where $X_c$ refers to any hadronic
system with a charm quark.
In addition, the moments of the lepton energy ($\El$)
and charmed hadron mass ($\mX$) distributions can be
calculated using HQE provided that the quantity in question
is integrated over a large region of the phase space to
ensure quark-hadron duality.

The expansion is performed in terms of $1/m_b$ and $\alpha_s(m_b)$,
with a mass scale parameter $\mu$ (typically 1\,GeV)
separating the short- and long-distance QCD effects.
While the former is perturbatively calculable given $m_b$, $m_c$,
and $\alpha_s$, the latter is not, and we are left with
parameters representing the expectation values of the non-perturbative
operators that appear in the expansion.
The measurement presented here determines $\Vcb$, $m_b$, $m_c$,
the branching fraction $\BR$, and four non-perturbative parameters
by a global fit to the lepton-energy and hadron-mass moments.

\section{Measurement}
We measured, in the BABAR data, the following eight moments:
\begin{equation}
M_0^{\ell} = \frac{\int{d\Gamma}}{\Gamma_B},\quad
M_1^{\ell} = \frac{\int{\El d\Gamma}}{\int{d\Gamma}},\quad
M_i^{\ell} = \frac{\int{(\El-M_1^{\ell})^i d\Gamma}}{\int{d\Gamma}}
\quad (i=2,3),
\end{equation}
\begin{equation}
M_i^{X} = \frac{\int{m_X^i d\Gamma}}{\int{d\Gamma}}
\quad (i=1,2,3,4),
\end{equation}
where $\Gamma_B$ is the total $B$ decay rate and
$d\Gamma$ is the differential $B\to X_c\ell\nu$ decay rate.
The integrations are done in the phase-space region in which
$\El$ is greater than an energy threshold $\Ecut$.

\subsection{Measurement of the lepton-energy moments}
The lepton-energy moments were measured~\cite{emoments}
using electrons found in the BABAR data that correspond
to $47.4\,\fbinv$ on the $\Upsilon(4S)$ resonance
and $9.1\,\fbinv$ below.
We selected the events containing two electrons:
one with a large center-of-mass momentum identified the event
as a likely $\Upsilon(4S)\to B\bar{B}$ decay;
the other was used to measure
the lepton energy spectrum.
The charge and angular correlations between the two electrons
were used to separate the conribution
of the primary $B\to X_c e\nu$ decays from the
other sources.
After correcting for the experimental
efficiencies and for the Bremsstrahlung in the detector
material,
four moments $M_i^{\ell}$ were calculated with $\Ecut$
varying between $0.6$ and $1.5\,\GeV$.
Corrections were applied for the difference between
the $\Upsilon(4S)$ and $B$ rest frames, for the effect
of the final state photon radiation, and for the
$B\to X_u\ell\nu$ decays.

\subsection{Measurement of the hadron-mass moments}
The hadron-mass moments were measured~\cite{hmoments}
in $81\,\fbinv$ of on-peak BABAR data.
We selected the events in which a $B$ meson was fully
reconstructed in a hadronic decay channel.
The remaining part of the event contains another $B$
meson, whose flavor and momentum are known by the
conservation laws.
In this recoil-$B$ sample, we searched for a lepton
accompanied by a missing 4-momentum compatible with a
neutrino.
The invariant mass of the hadronic system was determined
by a kinematic fit assuming the 4-momentum conservation,
the $B$ meson mass, and zero neutrino mass.
Residual downward bias in the measured mass, due mainly
to the undetected particles, was corrected using Monte-Carlo
simulation.
Four moments $M_i^X$ were calculated with $\Ecut$
varying between $0.9$ and $1.6\,\GeV$.

\subsection{Fit to the HQE prediction}
The HQE fit~\cite{hqefit} used the calcuation in the
kinetic mass scheme by Gambino and
Uraltsev~\cite{gambinouraltsev}.
It contains four non-perturbative paramters:
$\mu_{\pi}^2$, $\mu_G^2$, $\rho_D^3$, and $\rho_{LS}^3$.
They represent the expectation values of the kinetic,
chromomagentic, Darwin, and spin-orbit operators, respectively.

The eight moments, with varying $\Ecut$, provided 48 data points,
some of which are heavily correlated.
The full error matrix was calculated for both the statistical and
systematic experimental errors.
For the theoretical uncertainties, the dependencies on the
$\mu_{\pi,G}^2$ and $\rho_{D,LS}^3$ parameters were varied
by $\pm20\%$ and $\pm30\%$, respectively.
The theoretical errors were assumed to be fully correlated
among each moment at different $\Ecut$,
but uncorrelated between different moments.
These assumptions and the sizes of the theoretical uncertainties were
varied to confirm that the fit results were stable.

The result of the nominal fit is shown in Fig.~1.
\begin{figure}
\begin{center}\psfig{file=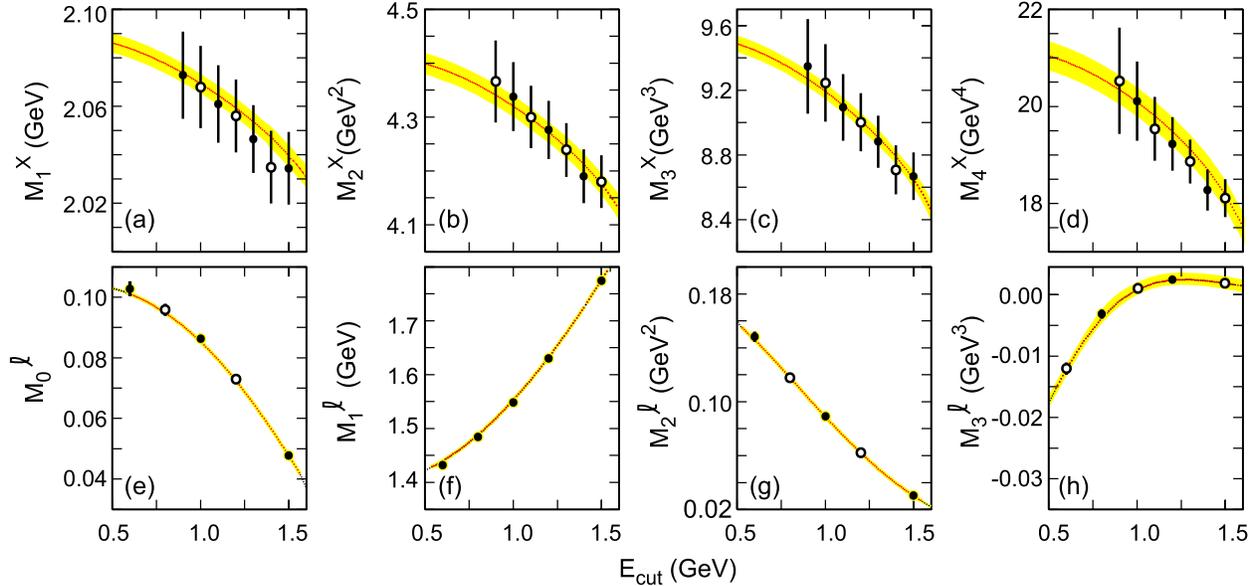,width=\textwidth}\end{center}
\caption{Results of the HQE fit. The points and the error bars are
the measured moments and their experimental errors. The curves and
the bands are the fit results and the theoretical errors.}
\end{figure}
The $\chi^2$ is 15 for 20 degrees of freedom.
Fits were repeated using $\El$ and $\mX$ moments separately.
The results, shown in Fig.~2, are consistent with the nominal fit.
\begin{figure}
\begin{center}\psfig{file=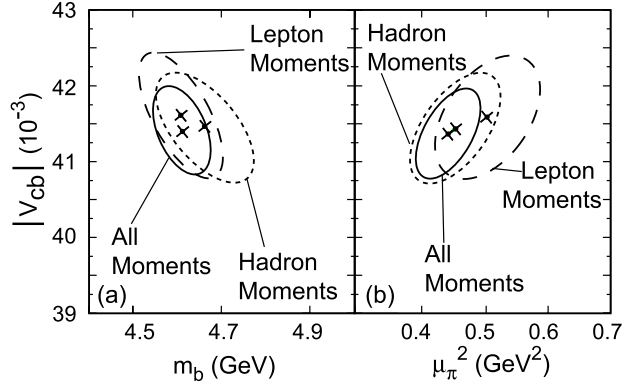,width=0.5\textwidth}\end{center}
\caption{One-sigma contours of the HQE fits using all moments,
lepton-energy moments only, and hadron-mass moments only.}
\end{figure}
We conclude from these observations that the HQE calculation
describes the experimental data very well.

The numerical results obtained from the fit are:
\begin{eqnarray*}
\Vcb &=& (41.4\pm0.4_{\mathrm{exp}}\pm0.4_{\mathrm{HQE}}\pm0.6_{\mathrm{th}})\times10^{-3},\\
\BR &=& (10.61\pm0.16_{\mathrm{exp}}\pm0.06_{\mathrm{HQE}})\%,\\
m_b^{\mathrm{kin}}(1\,\GeV) &=& (4.61\pm0.05_{\mathrm{exp}}\pm0.04_{\mathrm{HQE}}\pm0.02_{\mathrm{\alpha_s}})\,\GeV,\\
m_c^{\mathrm{kin}}(1\,\GeV) &=& (1.18\pm0.07_{\mathrm{exp}}\pm0.06_{\mathrm{HQE}}\pm0.02_{\mathrm{\alpha_s}})\,\GeV,\\
\mu_{\pi}^2 &=& (0.45\pm0.04_{\mathrm{exp}}\pm0.04_{\mathrm{HQE}}\pm0.01_{\mathrm{\alpha_s}})\,\GeV^2,\\
\mu_{G}^2 &=& (0.27\pm0.06_{\mathrm{exp}}\pm0.03_{\mathrm{HQE}}\pm0.02_{\mathrm{\alpha_s}})\,\GeV^2,\\
\rho_{D}^3 &=& (0.20\pm0.02_{\mathrm{exp}}\pm0.02_{\mathrm{HQE}}\pm0.00_{\mathrm{\alpha_s}})\,\GeV^3,\\
\rho_{LS}^3 &=& (-0.09\pm0.04_{\mathrm{exp}}\pm0.07_{\mathrm{HQE}}\pm0.01_{\mathrm{\alpha_s}})\,\GeV^3.
\end{eqnarray*}
The last error on $\Vcb$ accounts for the uncalculated theoretical
corrections to the semileptonic decay rate.
The kinetic quark masses obtained above can be translated into the
$\overline{\mathrm{MS}}$ masses:
\[
\overline{m}_b(\overline{m}_b) = 4.22\pm0.06\,\GeV,\quad
\overline{m}_c(\overline{m}_c) = 1.33\pm0.10\,\GeV.
\]
These results represent one of the most
accurate experimental determination of the heavy quark masses.

\section{Summary}
We determined $\Vcb$, $m_b$, $m_c$, $\BR$,
and four non-perturbative parameters by a global fit to the
lepton-energy and hadron-mass moments.
The HQE fit found an excellent agreement and consistency
between the theory and the data,
giving us confidence in the validity of the application.
Our fit used no external constraints on the HQE parameters,
an improvement over the previous HQE-based
measurements~\cite{previous} in which some or all the
parameters were either fixed or constrained.


\begin{thebibliography}{0}
\bibitem{emoments} B.~Aubert \textit{et al.} [BABAR Collaboration],
hep-ex/0403030, March 2004, to be published in Phys.\ Rev.\ D.

\bibitem{hmoments} B.~Aubert \textit{et al.} [BABAR Collaboration],
hep-ex/0403031, March 2004, to be published in Phys.\ Rev.\ D.

\bibitem{hqefit} B.~Aubert \textit{et al.} [BABAR Collaboration], 
hep-ex/0404017, April 2004, to be published in Phys.\ Rev.\ Lett.

\bibitem{gambinouraltsev} P.~Gambino and N.~Uraltsev,
Eur.\ Phys.\ J.\ \textbf{C34}, 181 (2004);
N.~Uraltsev, hep-ph/0403166, March 2004.

\bibitem{previous}
D.~Cronin-Hennessy \textit{et al.} [CLEO Collaboration],
Phys.\ Rev.\ Lett.\ \textbf{87}, 251808 (2001);
C.~W.\ Bauer, Z.~Ligeti, M.~Luke, and A.~V.\ Manohar,
Phys.\ Rev.\ \textbf{D67} 054012 (2003);
A.~H.\ Mahmood \textit{et al.} [CLEO Collaboration],
Phys.\ Rev.\ \textbf{D67} 072001 (2003); 
M.~Battaglia \textit{et al.} [DELPHI Collaboration],
Phys.\ Lett.\ \textbf{B556}, 41 (2003).

\end{thebibliography}
\end{document}